\documentstyle[prl,aps,multicol,epsf]{revtex}

\twocolumn
\begin{document}
\title{Unconventional geometric quantum computation}
\author{Shi-Liang Zhu$^{1,2}$
and Z. D. Wang$^{1,3}$
\thanks{zwang@hkucc.hku.hk}}
\address{$^1$Department of Physics, University of Hong Kong, Pokfulam Road,
Hong Kong, China\\
$^2$ Department of Physics,
South China Normal University, Guangzhou, China\\
$^3$ Department of Material Science and Engineering, University of
Science and Technology of China, Hefei, China
}
\address{\mbox{}}
\address{\parbox{14cm}{\rm \mbox{}\mbox{}
We propose a new class of unconventional geometric gates involving
nonzero dynamic phases, and elucidate that geometric quantum
computation can be implemented by using these gates. Comparing
with the conventional geometric gate operation, in which the
dynamic phase shift must be removed or avoided, the gates proposed here may be
operated more simply. We illustrate in detail that unconventional
nontrivial two-qubit geometric gates with built-in fault tolerant
geometric features can be implemented in real physical systems.
}}
\address{\mbox{}}
\address{\parbox{14cm}{\rm PACS numbers: 03.67.Lx, 03.65.Vf, 03.67.Pp}}
\maketitle


\newpage
\narrowtext



Quantum computation takes its power from superposition and
entanglement, which are two main features distinguishing the
quantum world from the classical world. But they are also very
fragile and may be destroyed easily by a process called
decoherence. The suppression of these decoherence effects in a
large-scalable quantum computer is essential for construction of
workable quantum logical devices. Quantum error-correcting
codes\cite{Shor1995} enable quantum computers to operate despite
some degree of decoherence and may make quantum computers
experimentally realizable, provided that the noise in individual
quantum gates is below a certain constant threshold. The recently
estimated threshold is that the individual gate infidelity should
be of the order $10^{-4}$ \cite{Steane2003}. In order for this
precision to be possible, quantum gates must be operated in a
built-in fault tolerant manner.

Apart from a decoherence-free scheme\cite{DFS}, 
a promising approach to achieve built-in fault tolerant quantum gates
is based on geometric phase
shifts\cite{Berry,Aharonov,Zhu_PRL2000}. A universal set of
quantum gates \cite{Lloyd} may be realized using geometric phase
shifts when the Hamiltonian of the qubit system changes along
suitable loops in a control
space\cite{Zanardi,Jones,Falci,Duan,Solinas,Wang,Zhu_PRL2002,Zhu_PRA2003}.
A quantum gate is expressed by a unitary evolution operator
$U(\{\gamma\})$, where the set $\{\gamma\}$ are phases acquired in
a particular evolution in realization of the gate, and usually
these phases consist of both  geometric $(\gamma^g)$ and a dynamic
$(\gamma^d)$ components\cite{Berry,Aharonov,Zhu_PRL2000}.
$U(\{\gamma\})$ is specified as a geometric gate if the phase
$\gamma$ in the gate-operation
 is a pure geometric one(i.e., with zero dynamic phase in the
evolution), and quantum computation implemented in this way is
referred to as geometric quantum computation (GQC) in a general
sense\cite{Zanardi,Jones,Falci,Duan,Solinas,Wang,Zhu_PRL2002,Zhu_PRA2003}.
GQC demands that logical gates in computing are realized by using
geometric phase shifts, so that it may have the inherently
fault-tolerant advantage due to the fact that the geometric phases
depend only on some global geometric features. Although this
property was doubted by some numerical calculations with certain
decohering mechanisms \cite{Nazir},  an analytical result showed
that geometric phases may be robust against
dephasing\cite{Carollo}. Several basic ideas of adiabatic GQC by
using NMR \cite{Jones}, superconducting nanocircuits \cite{Falci},
trapped ions \cite{Duan}, or semiconductor nanostructure
\cite{Solinas} were proposed, and the generalization to
nonadiabatic case was also
suggested\cite{Wang,Zhu_PRL2002,Zhu_PRA2003}.

 According to conventional wisdoms, a key point in GQC
\cite{Zanardi,Jones,Falci,Duan,Solinas,Wang,Zhu_PRL2002,Zhu_PRA2003}is
to remove/avoid the dynamic phase. One simple method is to choose the
dark states as qubit space, thus the dynamic phase is always zero
\cite{Duan}.  A more general method to cancel the dynamic phase is
the  so-called multi-loop scheme, i.e., let the evolution be
dragged by the Hamiltonian along several special closed loops,
then the dynamic phases accumulated in different loops may be
cancelled, with the geometric phases being
added\cite{Jones,Falci,Zhu_PRA2003}. These methods to cancel the
dynamic phase need subtle choice of the control parameters and/or
more operations than that needed in dynamic phase gates, and thus
may induce additional errors in the operations. On the other hand,
since the central idea of the GQC is that the phase accumulated in
the gate evolution has global geometric features, it is nature  to
ask whether we can design and implement a quantum gate with
geometric features but a nonzero/non-trivial dynamic phase.
Clearly, this kind of gate  differs from the conventional
geometric quantum gates addressed
before\cite{Zanardi,Jones,Falci,Duan,Solinas,Wang,Zhu_PRL2002,Zhu_PRA2003}
and is of significance in physical implementation of the built-in
fault tolerant quantum computation.

In this Letter, we not only answer the above important question
clearly for the first time, but also propose a new class of unconventional geometric
quantum gates,
 in which the total phase $\gamma$ consists of both a
 geometric component and a nonzero
dynamic one. Our novel and key idea is simply that despite its nonzero dynamic component,
 the total
phase is still dependent only on global geometric features if we
ensure that the dynamic phase $\gamma^d$ is proportional to the
geometric phase $\gamma^g$\cite{Berry,Aharonov,Zhu_PRL2000} as
\begin{equation}
\label{proportional} \gamma^d=\eta\gamma^g, \ \ (\eta \neq  0, \ -1),
\end{equation}
where $\eta$ is a proportional constant  independent on (or at
least some) parameters of the qubit system.
Eq.(\ref{proportional}) may be rewritten as
$\gamma=(1+\eta)\gamma^g$ with $\gamma$ the total phase
accumulated in the gate-operation, and the corresponding quantum
gate should possess global geometric features, which  we hereafter
specify as the unconventional geometric gate.
 These gates
would have advantages that conventional geometric gates have.
Comparing with conventional geometric gates, unconventional
geometric gates proposed here can simplify experimental
operations, since the additional operations required to cancel the
dynamic phase are not necessary in certain physical systems. In
the following, we illustrate in detail that unconventional
nontrivial two-qubit geometric gates with inherent fault tolerant
geometric features can be really implemented in physical
systems\cite{Cirac1995,Leibfried,Milburn,Sorensen2000}, and
specify the recently reported two-qubit phase gate\cite{Leibfried}
as an unconventional geometric gate proposed here.

 Let us consider a realistic physical system proposed quite earlier
in implementing quantum computers\cite{Cirac1995}.
In this system, two ions are confined
in a harmonic trap potential and interact with laser radiation.
 Two internal states of each ion
denoted by $|\downarrow\rangle$ or $|\uparrow\rangle$ represent
the qubit states. By choosing the laser beams appropriately, the
trap potential may excite a stretch mode with the frequency
$\omega_s$ when the ions are in the different internal state,
while nothing happened when they are in the same internal
state\cite{Leibfried}. If the internal states are in $|\downarrow
\uparrow\rangle$ or $|\uparrow\downarrow\rangle$, within the
rotating wave approximation,
 the Hamiltonian
of this system in the rotating frame reads ($\hbar=1$)
\begin{equation}
\label{Hamiltonian} H(t)=i\Omega_D (a^\dagger e^{-i\delta
t+i\phi_L}-a e^{i\delta t-i\phi_L}),
\end{equation}
where $a^\dagger$ and $a$ are the usual harmonic oscillator
raising and lowering operators, $\delta$ is the detuning, $\phi_L$
represents the phase of the driving field and
$\Omega_D=-(F_{0\downarrow}-F_{0\uparrow})z_{0s}/2$ with $z_{0s}$
being the spread of the ground state wave function of the stretch
mode. $F_{0\downarrow}$ ($F_{0\uparrow}$) is the dipole force
acting on the $|\downarrow\rangle$ ($|\uparrow\rangle$) state. The
quantum state $|\Psi\rangle$ under this force can be coherently
displaced in position-momentum phase space. It is clear that the
populations of the two ions would not change when the system is
governed by the Hamiltonian (\ref{Hamiltonian}), thus the
two-qubit gate achieved in the cyclic evolution should be a phase gate
described by
\begin{equation}
\label{Phase_gate}
U(\gamma)=\text{diag}[1,\exp(i\gamma_{\downarrow\uparrow}),
\exp(i\gamma_{\uparrow\downarrow}),1]
\end{equation}
in the computational basis $\{|\downarrow\downarrow\rangle,
|\downarrow\uparrow\rangle,|\uparrow\downarrow\rangle,
|\uparrow\uparrow\rangle \}$
($\gamma_{\downarrow\uparrow}=\gamma_{\uparrow\downarrow}=\gamma$)
\cite{XWang} .
The phase gate
(\ref{Phase_gate}) is a nontrivial two-qubit gate when $\gamma
\neq n\pi$ with $n$ an integer \cite{Lloyd}.

 To express explicitly the geometric and dynamic phases,
 we here employ the coherent-state path integral formulation
in the phase space to derive them. The phase change associated
with cyclic evolution in $[0,T]$ is defined by
$|\Psi(T)\rangle=\exp(i\gamma)|\Psi(0)\rangle$ with $\gamma$ a
real number. In order to evaluate $\gamma$, we may rewrite the
relation as $\exp(i\gamma)=\langle \Psi(0)|U(T,0)|\Psi(0)\rangle$
with the evolution operator being written as the standard
time-ordered product
\begin{equation}
\label{U_operator} U(T,0)=\hat{T}e^{-i\int_0^T
H(t)dt}=\prod_{n=1}^N e^{-iH(n)\epsilon}.
\end{equation}
Here $\hat{T}$ is the time
ordering operator, and $H(n)$ denotes the Hamiltonian at time
$t=t_n$. In the system we consider here, we may use the coherent
states $|\alpha_{\downarrow\uparrow}\rangle$, which are the
eigenstates of the destruction operator $a$ with eigenvalues
$\alpha_{\downarrow\uparrow}$. If $H(a^\dagger,a;t)$ is in
normally ordered (a more general case than that addressed here),
by inserting  $N$ resolutions of the identity $(1/\pi)\int
|\alpha_{\downarrow\uparrow}\rangle d^2\alpha_{\downarrow\uparrow}
| \alpha_{\downarrow\uparrow}\rangle=1$ into $U(T,0)$ with
$d^2\alpha_{\downarrow\uparrow}=d\mbox{Re}(\alpha_{\downarrow\uparrow})d\mbox{Im}(\alpha_{\downarrow\uparrow})$
 and $N \to \infty$, we
find that the propagator, defined as
$K(\alpha_{\downarrow\uparrow}(T);\alpha_{\downarrow\uparrow}(0))=\langle
\alpha_{\downarrow\uparrow}(0)|\alpha_{\downarrow\uparrow}(T)\rangle$,
is given by \cite{Hillery}
\begin{equation}
\label{Propagator}
K(\alpha_{\downarrow\uparrow}(T);\alpha_{\downarrow\uparrow}(0))
=\int
e^{i(\gamma^g+\gamma^d)}{\mathcal{D}}[\alpha_{\downarrow\uparrow}
(t)],
\end{equation}
where ${\mathcal{D}}[\alpha_{\downarrow\uparrow} (t)] \equiv
\lim_{N \to \infty} (1/\pi)^N \prod_{n=1}^{N}
d^2\alpha_{\downarrow\uparrow} (t_n)$,
\begin{equation}
\label{Phase_g}
 \gamma^g=\frac{i}{2}\int_0^T (\alpha_{\downarrow\uparrow}^\ast
\dot{\alpha}_{\downarrow\uparrow}-\dot{\alpha}_{\downarrow\uparrow}^\ast
\alpha_{\downarrow\uparrow})dt
\end{equation}
is just the geometric phase in a closed path in the phase
space\cite{Aharonov}, and
\begin{equation}
\label{Phase_d} \gamma^d=-\int_0^T
H(\alpha_{\downarrow\uparrow}^\ast,\alpha_{\downarrow\uparrow};t)dt
\end{equation}
is the dynamic phase with
$H(\alpha_{\downarrow\uparrow}^\ast,\alpha_{\downarrow\uparrow};t)=\langle
\alpha_{\downarrow\uparrow}|H(t)|\alpha_{\downarrow\uparrow}\rangle$
 \cite{Kuratsuji}. $\gamma^g$ in
  Eq.(\ref{Phase_g}) can also be expressed as
$(i/2)\oint [\alpha_{\downarrow\uparrow}^\ast
d\alpha_{\downarrow\uparrow}-\alpha_{\downarrow\uparrow}d\alpha_{\downarrow\uparrow}^\ast
]$, which is the area enclosed by the closed path of
$\alpha_{\downarrow\uparrow}(t)$. In the present system, we
have
\begin{equation} \label{alpha} \alpha_{\downarrow\uparrow}
(t)=i\frac{\Omega_D}{\delta}(e^{-i\delta t}-1)e^{i\phi_L},
\end{equation}
\begin{equation}
\label{H_path}
H(\alpha_{\downarrow\uparrow}^\ast,\alpha_{\downarrow\uparrow};t)=2\frac{\Omega_D^2}{\delta}[1-\cos(\delta
t)],
\end{equation}
under the condition that the initial state
$\alpha_{\downarrow\uparrow} (0)=0$. Therefore, the phases
accumulated in one cycle is found to satisfy Eq.(1)
\begin{equation}
\label{All_phases} \gamma^d  =-2\gamma^g =2\gamma  =2
\Phi_{\downarrow \uparrow}
\end{equation}
with $\Phi_{\downarrow \uparrow}=-2\pi (\Omega_D/\delta)^2$. Thus
a universal unconventional geometric gate described by Eq.(\ref{Phase_gate})
 can be
achieved once $\alpha_{\downarrow\uparrow} (t)$ forms a close path
in the gate operation. For instance, $\Phi_{\downarrow
\uparrow}=-\pi/2$ is obtained by choosing $|\Omega_D/\delta|=1/2$,
then $U(\pi/2)$ is a universal controlled $\pi-$phase gate after
rotating $-\pi/2$ on the $|\uparrow\rangle$ states.

At this stage, it is worth pointing out that the conventional
geometric phase gate is unreachable in this system since only a
nontrivial $\gamma^g=\pm \pi$ can be obtained under the condition
of trivial dynamic phase $(\gamma^d=2\pi\times \mbox{integer})$.
But $U(\gamma)$ is not a universal gate when $\gamma=\pm
\pi$\cite{Lloyd}. However, most intriguingly, the total phase is
exactly equal to  the minus geometric phase, namely,
 $U(\gamma)$ achieved here is a unconventional geometric logical gate with
$\eta=-2$ proposed before. Interestingly, the proportional constant in this
example is indeed independent of any parameters in the system,
such as the speed of the gate, the detuning, the phase and the
density of the laser beams used {\it etc.}. Consequently,
the phase $\gamma$ in the gate (\ref{Phase_gate})  has really all
the features depend only on the geometry. Remarkably, the phase
gate addressed here was experimentally demonstrated
very recently\cite{Leibfried}, and the high fidelity of the two-qubit
phase gate achieved in the reported experiment benefits from its
geometric features: the phase is determined only by the path area,
not on the exact starting state distributions, path shape,
orientation in phase space, or the passage rate to traverse the
closed path\cite{Leibfried}. All these features are global
properties which motivated one to study the conventional
GQC\cite{Zanardi,Jones,Falci,Duan,Solinas,Wang,Zhu_PRL2002,Zhu_PRA2003}.
To our knowledge, a
conventional geometric quantum gate has not been achieved
experimentally, though the conditional geometric phase was
observed in Ref.\cite{Jones},

Also intriguingly, we find that Eq. (\ref{All_phases}) is valid
even in noncyclic cases, which has close relevance to the robustness of
the (cyclic phase) gate against the small noncyclic perturbations.
When a quantum system evolves from an initial state
$|\Psi (0)\rangle$ to a final state $|\Psi (t)\rangle$ with
 $ \langle\Psi (0)|\Psi
(t)\rangle=e^{i\gamma}|\langle\Psi (0)|\Psi (t)\rangle|$,
 $\gamma$ is  specified as the total phase and  the
 noncyclic geometric
phase can be defined as $\gamma^g=\gamma-\gamma^d$, where
$\gamma^d=-\int_0^t \langle \Psi (t^\prime)|H(t^\prime)|\Psi
(t^\prime)\rangle dt^\prime$ is the dynamic phase
\cite{Aharonov,Zhu_PRL2000}. In the present system, the wave
function $|\Psi (t) \rangle$ at time $t$ is $|\Psi (t) \rangle =
U_t |\psi (0)\rangle$, and the evolution operator $U_t$ can be
found as
\begin{eqnarray}
\nonumber
U_t &=& \hat{T} e^{-i\int_0^t H(t^\prime)dt^\prime}  \\
\label{Magnus}
     &=& e^{ -i\int_0^t H(t^\prime)dt^\prime
     -\frac{1}{2} \int_0^t dt_2 \int_0^{t_2} [H(t_2),H(t_1)] dt_1+\cdot\cdot\cdot } \\
\nonumber
     &=& e^{i\Phi_{\downarrow \uparrow} (t)}
     D(\alpha_{\downarrow\uparrow}),
\end{eqnarray}
where $ \Phi_{\downarrow \uparrow} (t) = (\Omega_D/\delta)^2
[sin(\delta t)-\delta t]$ and $\alpha_{\downarrow \uparrow} (t)$
is given by Eq. (\ref{alpha}). The commutator of the Hamiltonian
(\ref{Hamiltonian}) at different time is a number, not an
operator. Then the last equation is exactly derived by expanding
the magnus' formula [Eq. (\ref{Magnus})] \cite{Magnus54} to the
second term, since the higher-order terms in the expansion vanish.
Then it is straightforward to derive $\gamma
(t)=\Phi_{\downarrow\uparrow} (t)$ at any time $t$. On the other
hand, the dynamic phase accumulated during $[0,t]$ can be obtained
explicitly $ \gamma^d (t) = -\int_0^t \langle n| D^\dagger
(\alpha_{\downarrow\uparrow})H(t^\prime)
D(\alpha_{\downarrow\uparrow}) |n\rangle dt^\prime = 2
\Phi_{\downarrow \uparrow} (t)$, where $|n\rangle$ is an
eigenstate in the Fock space. Therefore, we conclude that Eq.
(\ref{All_phases}) is valid at any time. Because of this very
special property, the total phase still depends only on the
geometric features even in the presence of a slight deviation of
the period T and thus the illustrated geometric gate is also
insensitive to the error in controlling the cyclic time in this
respect\cite{note}, which is an extra advantage of this kind of
gate and is believed to be one of factors leading to high fidelity of the
unconventional phase gate reported experimentally\cite{Leibfried}.


We now turn to another interesting example\cite{Milburn}.  In ions trapped
quantum computer model, the Hamiltonian for ions interacting with
the vibrational mode can be controlled by using different kinds of
Raman laser pulses. In the case of two ions with each driven by
identical Ramman lasers, the system  may be described by a
special case of the interaction Hamiltonian given by
\begin{equation}
\label{Hamiltonian2}
H(t)=-i[f(t)a^\dagger-f^\ast (t) a]
\hat{J}_z,
\end{equation}
where $\hat{J}_z=\sigma_z^{(1)}+\sigma_z^{(2)}$ with
$\sigma_z^{(j)}$ being the $z$-component Pauli matrix for the
$j$th ion is the collective spin operator. The conditional phase
gate in the system has been proposed by using specific four pulse
sequence \cite{Milburn}. We suggest a more general gate achieved
by this Hamiltonian than that addressed there. The gate governed
by the Hamiltonian (\ref{Hamiltonian2}) is clearly a phase gate
$U(\{ \gamma
\})=\text{diag}[\exp(i\gamma_{\downarrow\downarrow}),\exp(i\gamma_{\downarrow\uparrow}),
\exp(i\gamma_{\uparrow\downarrow}),\exp(i\gamma_{\uparrow\uparrow})]$,
since  this Hamiltonian would not lead the spins to flip in the
computational basis. Denoting $\beta_{jl}$ ($j,l=\downarrow,\
\mbox {or} \uparrow$) as the eigenvalues for $\hat{J}_z$ in this
basis, it is straightforward to find that
\begin{eqnarray*}
 D^\dagger (\beta_{jl}) a^\dagger \beta_{jl}
D(\beta_{jl}) &=& \beta_{jl}^2 (a^\dagger+\alpha^\ast(t)),   \\
 D^\dagger (\beta_{jl}) a \beta_{jl} D
(\beta_{jl})&=&\beta_{jl}^2 (a+\alpha(t)),
\end{eqnarray*}
where $D(\beta_{jl})=\exp \{[\alpha (t) a^\dagger-\alpha^\ast (t)
a]\beta_{jl}\}$ with $\alpha (t)=-\int_0^t f(t^\prime)dt^\prime$.
Then we have,
\begin{eqnarray*}
 H(\alpha^\ast,\alpha;t)&=&\langle n| D^\dagger
(\beta_{jl}) H(t) D(\beta_{jl}) |n\rangle\\
 \label{H2}
&=&-i\beta_{jl}^2 [f(t)\alpha^\ast (t)-f^\ast (t)\alpha (t)].
\end{eqnarray*}
Substituting this result into Eq.(\ref{Phase_d}), the dynamic
phase is given by
\begin{equation}
\label{Pd2}
 \gamma_{jl}^d(\tau)=2\beta^2_{jl}\gamma^0 (\tau),
\end{equation}
with $\gamma^0 (\tau)=(1/2)\int_0^\tau [\alpha^\ast(t)f(t)-\alpha
(t)f^\ast (t)]dt.$
 The geometric
phase is then found to be
\begin{equation}
\label{Pg2} \gamma^g_{jl} (\tau)=-\beta_{jl}^2 \gamma^0 (\tau).
\end{equation}
Comparing Eq.(\ref{Pd2}) with Eq. (\ref{Pg2}), we have
\begin{equation}
\label{All_phases2} \gamma^d_{jl} (\tau) =-2\gamma^g_{jl}
(\tau)=2\gamma_{jl} (\tau).
\end{equation}
Thus a universal phase gate $U(\{\gamma\})$ may also be realized
if $\alpha (t)$ forms a closed path, noting that $U(\{\gamma\})$
is nontrivial under the condition
$\gamma_{\downarrow\downarrow}+\gamma_{\uparrow\uparrow} \neq
\gamma_{\downarrow\uparrow}+\gamma_{\uparrow\downarrow}$
$(\mbox{mod}\ 2\pi)$.


Similarly, by appropriately choosing laser beams, the ions in a
Paul trap may be described by the Hamiltonian given by
$$
\label{Hamiltonian3} H(t)=-i[f(t)a^\dagger-f^\ast (t) a]
\hat{J}_y,
$$
with $\hat{J}_y=\sigma^{(1)}_y+\sigma^{(2)}_y$. Comparing with the
Hamiltonian (\ref{Hamiltonian2}),  only  a basis changes from
$\hat{J}_z$ to $\hat{J}_y$. Using a similar method, we find a gate
given by $U(\tau)=\exp(-i\gamma (\tau)\hat{J}_y^2)$ with $\gamma
(\tau)=\gamma^0 (\tau)$\cite{Sorensen2000} , and also have
$\gamma^d (\tau) =-2\gamma^g (\tau)=2\gamma (\tau)$.

 Clearly, the quantum gates demonstrated above
are just the unconventional geometric gates with a
parameter-independent proportional constant. Therefore, they not
only possess all geometric advantages that conventional geometric
gates have but also are independent of initial states in the
system, enabling one to reach the high fidelity.
Nevertheless, we should note that
the uncertainty of the phase in a general unconventional geometric
quantum gate
 comes from
two factors: fluctuations due to the conventional geometric
phase term and the $\eta$ term. Generally speaking,
an unconventional geometric gate is robust to the fluctuations or
perturbations from the parameters which $\eta$ (and $\gamma^g$) is
independent of. This is the reason why $\eta$ is  required to be
independent on at least some parameters of the qubit system;  a
perfect unconventional geometric gate is just the example
illustrated above: $\eta$ is independent on all parameters of the
system.

 To conclude, we have proposed a new class of unconventional
geometric quantum gates. Comparing with conventional GQC, our
proposal may  simplify experimental operations, because additional
operations to remove/avoid the dynamic phase are no longer
required. Apart from the above-addressed systems related to the
harmonic oscillators, it is  of great significance to design and
to implement this class of unconventional geometric gates in other
physical systems.

We thank Paolo Zanardi for his critical reading of this paper
and useful discussions as well as suggestions.
This work was supported by the RGC grant of Hong Kong under No.
HKU7114/02P and a URC fund of HKU. S.L.Z was supported in part by
the NSF of Guangdong under Grant No. 021088 and the NNSF of China
under Grant No. 10204008.


\begin{thebibliography}{99}
\bibitem{note} Note that in the presence of small noncyclic perturbations
the gate operator (\ref{Phase_gate}) may be modified as
$\text{diag}[1,\nu\exp(i\gamma), \nu\exp(i\gamma),1]$ with $\nu$
($\in [0,1]$) the visibility
being slightly less than unit, which may lead to a weak
decoherence effect; but it is negligible even inevitable.
Quantitatively, for a small deviation $\Delta T$ from the period $T$,
 $\nu$ is estimated to be
$1-(2\pi \Omega_D/\sqrt{2} \delta)^2 (\Delta T/T)^2$ and the deviation
of $\gamma$ is about $-2\pi(2\pi \Omega_D/\sqrt{6} \delta)^2 (\Delta T/T)^3$.
\end{thebibliography}
\end{document}